\documentclass[10pt,aps,pre,twocolumn,superscriptaddress,nobibnotes,nodoi,noeprint]{revtex4-2}

\usepackage{graphicx}
\usepackage{amsmath,physics,bm,float}
\usepackage{soul}
\usepackage[colorlinks=true,citecolor=blue,urlcolor=blue,linkcolor=blue]{hyperref}
\usepackage{amssymb}
\usepackage{natbib}
\usepackage{upgreek}
\usepackage{mathtools}

\graphicspath{{Figures/}}

\usepackage[mathlines]{lineno}
\let\oldequation\equation\let\oldendequation\endequation
\renewenvironment{equation}{\linenomathNonumbers\oldequation}{\oldendequation\endlinenomath}
\let\oldalign\align\let\oldendalign\endalign
\renewenvironment{align}{\linenomathNonumbers\oldalign}{\oldendalign\endlinenomath}

\DeclareMathOperator*{\sign}{sign}

\def\CC{{C\nolinebreak[4]\hspace{-.05em}\raisebox{.4ex}{\tiny\bf ++}}}

\begin{document}

\title{Temperature overshooting in the Mpemba effect of frictional active matter}

\author{Alexander P.\ Antonov}
\email{alexander.antonov@hhu.de}
\affiliation{
Institut f{\"u}r Theoretische Physik II: Weiche Materie,
Heinrich-Heine-Universit{\"a}t D{\"u}sseldorf, 
D-40225 D{\"u}sseldorf, 
Germany}

\author{Hartmut L{\"o}wen}
\affiliation{
Institut f{\"u}r Theoretische Physik II: Weiche Materie,
Heinrich-Heine-Universit{\"a}t D{\"u}sseldorf, 
D-40225 D{\"u}sseldorf, 
Germany}

\date{\today}
\begin{abstract}
The traditional Mpemba effect refers to an anomalous cooling
phenomenon when an initial hotter system cools down faster than an
initial warm system. Such counterintuitive behavior has been confirmed
and explored across phase transitions in condensed matter systems and
also for colloidal particles exposed to a double-well potential. Here we
predict a frictional Mpemba effect for a macroscopic body moving
actively on a surface governed by Coulomb (dry) friction.
For an initial high temperature, relaxation towards a
cold state occurs much faster than that for an intermediate initial
temperature, due to a large temperature overshooting in the latter case.
This frictional Mpemba effect can be exploited to steer the
motion of robots and granules.
\end{abstract}

\maketitle

\section{Introduction}

The Mpemba effect -- first reported in antiquity in Aristotle's \textit{Meteorologica} \cite{aristotle} and later famously associated with Erasto Mpemba himself, who observed it while freezing ice cream \cite{mpemba1969cool} -- refers to the counterintuitive phenomenon in which a system initially at a higher temperature reaches its steady state faster than an identical system starting from a lower temperature when both are quenched into the same cold environment. Controversial by nature, this effect has attracted broad attention and has been observed since then in a wide range of systems. Beyond the canonical case of water freezing, notable instances include but are not limited to colloidal suspensions \cite{kumar2020exponentially, kumar2022anomalous, malhotra2024double} and quantum systems \cite{Chatterjee2023, Nava/Egger:2024, Joshi2024, Moroder/etal:2024, liu2025symmetry}. Nevertheless, despite its consistent phenomenological signature, the underlying mechanisms appear to be system-specific and remain the subject of ongoing debate \cite{bechhoefer2021fresh}, lacking even a universal heuristic -- let alone a theoretical -- explanation.

One of the leading theoretical explanations for the Mpemba effect invokes a complex energy landscape featuring multiple metastable states \cite{lu2017nonequilibrium, Klich/etal:2019}. For simplicity, such landscapes are often visualized as effective potentials with just two minima -- a local and a global one \cite{kumar2020exponentially, bechhoefer2021fresh}. In this framework, a system initialized at a higher temperature is less likely to be found near the local minimum, and thus less likely to become transiently trapped there. Conversely, a system starting from a lower temperature is more probable to occupy a local minimum, resulting in a delayed relaxation to the global one.

While this explanation provides an intuitive understanding of Mpemba-like behavior, it must be adapted for systems where no well-defined energy landscape exists, such as those governed by non-conservative forces or subject to strong nonequilibrium driving. A prominent class of such systems is active matter, characterized by the persistent injection of energy from the environment or internal sources into directed (self-propelled) motion \cite{marchetti2013hydrodynamics, bechinger2016active}. In these cases, even the notion of temperature becomes ambiguous, and a fundamental question arises: \textit{how} \cite{hecht2024define} should temperature be defined in systems far from equilibrium? 

\begin{figure*}[htp!]
\includegraphics[width=\linewidth]{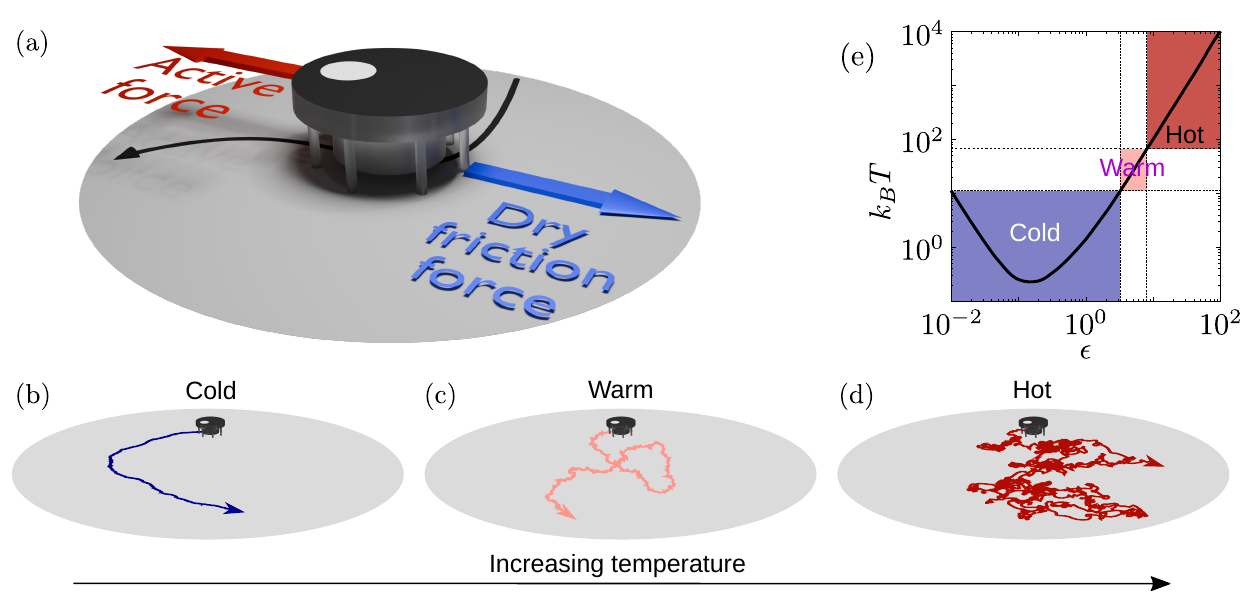}
\caption{(a) An illustrative example of active granular particle (vibrating robot), subjected to dry friction due to the solid-solid contact between the surface and its legs. (b)-(d) Trajectories sketched to illustrate different temperatures. The motion occurs due to an interplay between active force and dry friction force, while the temperature regulates the impact of the fluctuations, resulting in the gradual trajectories for cold (b) temperatures and in fluctuating trajectories when increasing the temperature for warm (c) and hot (d). (e) Dependence of the system temperature $T$ on the noise parameter $\epsilon$. Colored rectangle projections mark the cold, warm (intermediate), and hot states, highlighting the associated ranges of temperature $T$ and noise strength $\epsilon$.}
\label{fig:illustration}
\end{figure*}

In this study, we report the observation of the Mpemba effect in active systems with dry friction, such as active granular matter composed of vibrating robots~\cite{kumar2014flocking, Agrawal2020, baconnier2022selective, Chor/etal:2023}, where particles experience solid-solid contact with the frictional surface (Fig.~\ref{fig:illustration}(a)). In contrast to colloidal systems -- including active colloids \cite{Schwarzendahl/Lowen:2022} -- where the Mpemba effect emerges under velocity-dependent Stokes friction and confinement by an external potential, the system studied here evolves without any external potential and is instead subject to velocity-independent Coulomb, or dry friction~\cite{deGennes:2005, Romanczuk2012}. Therefore, we refer to it as a \textit{frictional Mpemba effect}. The intrinsic interplay between dry friction and active self-propulsion gives rise to several distinct behaviors, observed both theoretically and experimentally, including a continuous switching between diffusive and accelerated motion \cite{antonov2024inertial} and a frictional-induced phase separation mechanism \cite{antonov2025self}. Notably, the steady-state velocity distribution in such system deviates significantly from the Gaussian (Maxwell-Boltzmann) form, which is often considered a prerequisite for observing the Mpemba effect, e.g.\ in granular fluids with non-elastic collisions \cite{Lasanta/etal:2017}. The Mpemba effect observed in our study arises due to an overshooting phenomenon, in which the system temporarily reaches the temperatures below that of the steady state. This overshooting has been identified in other systems as a possible key mechanism underlying anomalous thermal relaxation that facilitates the occurrence of Mpemba-like behavior \cite{Klich/etal:2019, Megias/etal:2022, yu2025symmetry}.

\section{Model}

Here we consider a minimal model to study an inertial active particle with mass $m$ subject to a dry friction force $\sigma(v)= \Delta_C\sign(v)$, where $\Delta_C$ is the dry friction coefficient. The dynamics is described by a one-dimensional Langevin equation for the particle velocity $v(t)=\dot{x}(t)$,
\begin{equation}
m\dot{v}(t) = -\sigma(v(t)) + \sqrt{2K}\xi(t) + n(t)f,
\label{eq:velocity}
\end{equation}
where $t$ is time, $\xi(t)$ is a Gaussian white noise process with zero mean and correlations $\langle\xi(t)\xi(t')\rangle = \delta(t'-t)$, $K$ is the white noise strength, the time-dependent term $n(t) f$ is the active force with activity amplitude $f$ and the stochastic process $n(t)$ chosen as an Ornstein-Uhlenbeck process:
\begin{equation}
    \dot{n}(t) =-\frac{n(t)}{\tau} + \sqrt{\frac{2}{\tau}}\eta(t).
    \label{eq:activity}
\end{equation}
Here $\tau$ is the persistence time, and $\eta(t)$ represents Gaussian white noise with zero mean and correlations $\langle\eta(t)\eta(t')\rangle = \delta(t'-t)$. This choice of the active force is therefore typically referred to in literature as active Ornstein-Uhlenbeck particle dynamics \cite{szamel2014self, maggi2014generalized, caprini2018active, martin2021statistical, PhysRevLett.129.048002}. In this study, we focus on the overall impact of stochastic contributions, both white and Ornstein-Uhlenbeck noise. Since these noise sources can stem from related physical mechanisms \cite{kumar2014flocking}, it is natural to regulate the temperature through a single composite parameter that reflects their combined effect.

In what follows, we set $\Delta_C$, $\sqrt{\tau K}/\Delta_C$, $\tau K/m\Delta_C$ as units of force, time, and length, respectively. 
Under this rescaling, the behavior of the system reads:
\begin{subequations}
\begin{eqnarray}
    \dot{v}(t) & = & -\sign(v(t)) + n(t)f_0 + \sqrt{2\epsilon}\,\xi(t),\label{eq:dry} \\
    \dot{n}(t) & = & -\epsilon{n(t)} + \sqrt{2\epsilon}\,\eta(t).\label{eq:OU}
\end{eqnarray}
\label{eq:dynamics}
\end{subequations}
It depends solely on two key dimensionless quantities: the reduced activity amplitude $f_0=f /\Delta_C$, which quantifies the strength of the active force relative to dry friction and is fixed at $f_0 = 0.5$ throughout this study, and the noise parameter $\epsilon=\sqrt{K/\tau}/\Delta_C$ which
controls both sources of noise in the system -- the white noise with strength $K$ and the active noise (Ornstein-Uhlenbeck) characterized by the persistence time $\tau$.\ Through its presence in the stochastic contributions of Eq.~\eqref{eq:dynamics}, $\epsilon$ thus governs the overall impact of fluctuations on the particle dynamics and thus determines the temperature of the system. In what follows, we adopt the kinetic temperature, defined as 
\begin{equation}
T = \frac{1}{2}\frac{m\langle v^2 \rangle}{k_B},
\end{equation}
as our measure of the system's temperature.

In the low temperature regime sketched in Fig.~\ref{fig:illustration}(b), the dynamics are gradual, while increasing the temperature in Fig.~\ref{fig:illustration}(c)–(d) amplifies the role of fluctuations, resulting in progressively noisier and more entangled trajectories. In Fig.~\ref{fig:illustration}(e), we demonstrate how temperature can be controlled by the noise parameter $\epsilon$.

\begin{figure}[hb!]
		\includegraphics[width=\columnwidth]{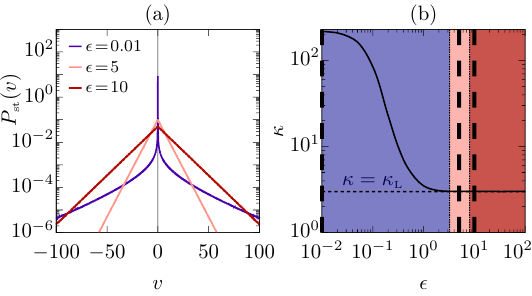}
		\caption{(a) Steady-state velocity probability distribution $P_{\rm st}(v)$ for various noise parameters $\epsilon$ and fixed activity amplitude $f_0 = 0.5$. (b) Kurtosis of the steady-state distribution as a function of noise parameter $\epsilon$ at a fixed activity amplitude $f_0 = 0.5$. Color phases correspond to those depicted in Fig.~\ref{fig:illustration}(e), and the dashed vertical lines correspond to the noise parameters of density profiles shown in panel (a).
        }
		\label{fig:steady-state}
\end{figure}

In case of dry friction, the particle does not move until the force acting on it exceeds the threshold value (which is $\Delta_C$, or 1 in our dimensionless units) of the friction force. Thus, the particle's motion can be initiated by either Brownian noise or an active force. We note that the active force contribution to the particle motion is unaffected by temperature, as the steady-state distribution of the stochastic process $n$, $P_{\rm st}(n) = e^{-n^2/2}/\sqrt{2\pi}$ is independent of $\epsilon$. However, the probability that the active force exceeds the threshold, 
\begin{equation}
\mathcal{P}(n f_0 \ge 1) = \int_{1/f_0}^{\infty} n\, P_{\rm st}(n) \, dn = \sqrt{\frac{1}{2\pi}}\,e^{-\frac{1}{2f_0^2}},
\end{equation}
remains small -- but non-negligible -- due to our specific choice of the activity amplitude $f_0 = 0.5$. 

At high noise parameter $\epsilon$, the motion is predominantly driven by Brownian noise, which dominates the activity term. In this noise-dominated regime ($\epsilon \gg f_0$), the dynamics \eqref{eq:dry} reduces to
\begin{subequations}
    \begin{equation}
        \dot{v}(t) = -\sign(v(t))+ \sqrt{2\epsilon}\,\xi(t),
    \end{equation}
    with the associated Fokker-Planck equation \cite{Risken1988}
    \begin{equation}
        \partial_t P(v, t) = \partial_v\left[\sign(v) P(v,t)\right] + \epsilon \partial_v^2 P(v,t).
    \end{equation}
\end{subequations}
Here, we have neglected the active term $n(t)f_0$, so this is an approximate description valid when $\epsilon \gg f_0$. The resulting steady-state velocity distribution reads:
\begin{equation}
P_{\rm st}(v; \epsilon) = \frac{1}{2\epsilon} e^{-\frac{|v|}{\epsilon}},
\label{eq:Laplacian}
\end{equation}
which exhibits a V-shaped profile in log-scale (Fig.~\ref{fig:steady-state}) and is characteristic for Brownian particles subjected to dry friction \cite{deGennes:2005, Touchette/etal:2010}. In contrast, at low $\epsilon$ the Brownian noise is suppressed by dry friction, and motion is only possible due to rare yet contributing fluctuations in the active force leading to accelerated motion.\ Such periods of accelerated motion, albeit rare, are lasting much longer due to the higher persistence of the Ornstein-Uhlenbeck process \eqref{eq:OU} at low noise parameter $\epsilon$, therefore leading to heavy distribution tails and a sharp peak at zero in the steady-state velocity distribution $P_{\rm st}(v; \epsilon)$ (Fig.~\ref{fig:steady-state}(a)). The kurtosis $\kappa = \langle v^4\rangle/(\langle v^2\rangle)^2 - 3$, shown in Fig.~\ref{fig:steady-state}(b), illustrates the transition from activity-dominated dynamics at low $\epsilon$ to noise-dominated behavior at high $\epsilon$. As noise increases, $\kappa$ decreases from large values, reflecting heavy-tailed velocity distributions, and approaches the kurtosis of a Laplacian distribution $\kappa_{\textrm{\tiny{L}}} = 3$. This is intrinsically different from most previously reported Mpemba-like effects caused by non-Gaussian velocity distributions, which occurred in the low-kurtosis regime \cite{Lasanta/etal:2017, santos2020mpemba}. 

\begin{figure*}[htp!]
        \includegraphics[width=0.95\linewidth]{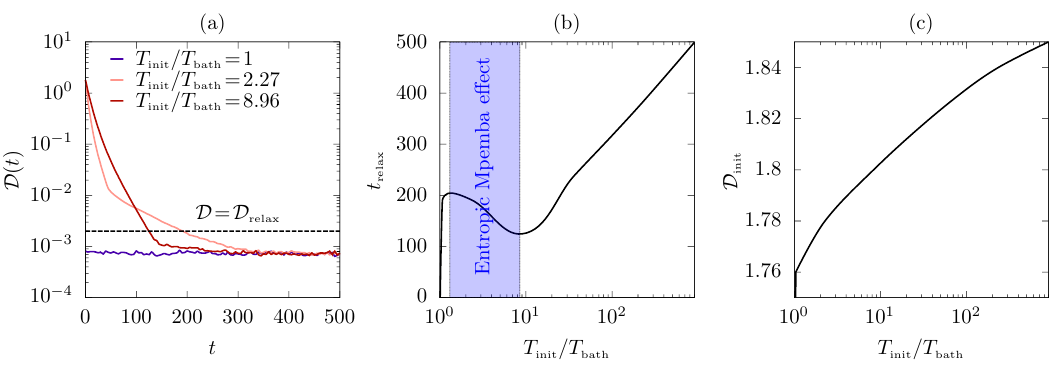}
		\caption{(a) Relaxation dynamics of the distance measure $\mathcal{D}(t)$.\ (b) Relaxation time in the temperature range $T_{\rm init}/T_{\rm bath} \in [1, 900]$.\ The blue rectangle indicates the temperature range where the entropic Mpemba effect (a reduction in the distance relaxation time) is observed.\ (c) Monotonic dependence of $\mathcal{D}_{\rm init}$ on $T_{\rm init}$ for the parameters employed in the numerical simulations.
        }
		\label{fig:Mpemba}
\end{figure*}

\begin{figure}[hb!]
        \includegraphics[width=\columnwidth]{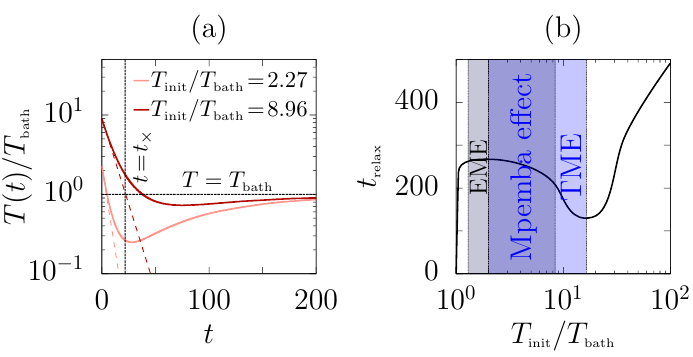}
		\caption{(a) Time evolution of system temperatures for warm and hot initial temperatures.\ (b) Relaxation time determined from the temperature evolution \eqref{eq:temp-relax}.\ The dark-blue rectangle (Mpemba effect) indicates the temperature range where both the thermal Mpemba effect (a reduction in the temperature relaxation time) and entropic Mpemba effects are observed; the light-blue rectangle (TME) indicates the temperature range where only the thermal Mpemba effect occurs, while the gray rectangle (EME) marks the range where only the entropic Mpemba effect is present.
        }
		\label{fig:Mpemba_temp}
\end{figure}

\section{Results}

To demonstrate the Mpemba effect, we consider cooling from initial warm and hot temperatures $T_{\rm init}$ set by $\epsilon_{\rm init}$, down to the bath temperature $T_{\rm bath} < T_{\rm init}$ set by $\epsilon_{\rm bath}$. First, we analyze the \textit{entropic Mpemba effect (EME) by comparing} how fast the probability distribution $P(v, t)$, initially distributed at $P(v, t=0) = P_{\rm st}(v; \epsilon_{\rm init})$ relaxes towards the steady-state distribution of the bath $P_{\rm st}(v; \epsilon_{\rm bath})$. The probability distributions $P(v,t)$ are obtained via simulations of the Langevin dynamics \eqref{eq:dynamics} for $\epsilon=\epsilon_{\rm bath}$ with $P(x,t) = \langle \delta(x-x(t))\rangle$, where $\langle \ldots \rangle$ denotes an ensemble average. Dynamics~\eqref{eq:dynamics} are simulated using the Euler-Maryama scheme, with timestep $dt = 10^{-3}$ and spatial resolution $dx=0.1$. All numerical results correspond to ensemble averages over $10^8$ independent
realizations. The full implementation and simulation details are available at \footnote{The \CC\ implementation is available from GitHub repository, \href{https://github.com/apantonov/dry-active}{https://github.com/apantonov/dry-active}}. To quantify the emergence of the EME, we define the distance measure as \cite{Schwarzendahl/Lowen:2022}:
\begin{subequations}
\label{eq:distance-relax}
\begin{equation}
    \mathcal{D}(t) = \int_{-\infty}^{\infty} dv \left| P(v,t) - P_{\rm st}(v; \epsilon_{\rm bath}) \right|,
\end{equation}
and the relaxation time $t_{\rm relax}$ is defined as the moment when the distance reaches 
\begin{equation}
\mathcal{D}(t_{\rm relax}) = \mathcal{D}_{\rm relax}.
\end{equation}
\end{subequations}
In our numerical simulations, we use $\mathcal{D}_{\rm relax} = 2 \times 10^{-3}$ as comparable to the noise level. Here, $\mathcal{D}_{\rm relax}$ is not a fundamental constant but a numerically determined threshold, chosen based on simulations of a system initially at $T_{\rm init} = T_{\rm bath}$. It is used only to indicate when the system has effectively reached the stationary distribution within the limits of numerical accuracy.

The distance evolution shown in Fig.~\ref{fig:Mpemba}(a) demonstrates a strong EME, since the initially hotter system reaches the steady state first despite starting at a greater distance from the bath distribution than the initially colder systems. The decrease in cooling time with increasing initial bath temperature, shown in Fig.~\ref{fig:Mpemba}(b), reveals the temperature range in which the Mpemba effect occurs. We note that, in general, an ordering in the temperatures,
$T_{{\rm init},1} > T_{{\rm init},2}$, corresponding to two distinct initial conditions, does not necessarily imply the same ordering
in the initial entropic distances,
$\mathcal{D}_{{\rm init},1} > \mathcal{D}_{{\rm init},2}$, where $\mathcal{D}_{\rm init} \equiv \mathcal{D}(t=0)$. Indeed, when considering entropic distances, higher-order moments of the steady-state velocity distribution, i.e.\ beyond the second moment associated with the kinetic temperature, may contribute~\cite{Megias/etal:2022}.
In the present work, however, we have explicitly verified numerically that 
$\mathcal{D}_{\rm init}$ is a monotonic function of $T_{\rm init}$
within the entire range of explored parameters, as illustrated in  Fig.~\ref{fig:Mpemba}(c).

But what are the factors that give rise to the Mpemba effect?\ To gain a better understanding of this phenomenon, we track the time evolution of the system (kinetic) temperature. At the first stages of the relaxation process depicted in Fig.~\ref{fig:Mpemba_temp}(a), the system is being slowed down (i.e.,\ cooled in terms of temperature) by the dry friction. To quantify this stage, we consider the evolution of the kinetic temperature, which characterizes the mean kinetic energy of the system.\\
In contrast to velocity-dependent friction types such as fluid friction, dry friction does not depend on velocity. Therefore, the cooling efficiency remains independent of the system velocity, unlike in the case of fluid friction, where faster-moving objects are slowed down (cooled) more efficiently than slower ones. In the considered bath regime, the white noise contribution is negligible, and activity becomes relevant only at later times due to the changed protocol (with the mean activation time $t_{\times} \approx 1/\mathcal{P}(nf_0 \ge 1)$ indicated by the dashed vertical line in Fig.~\ref{fig:Mpemba_temp}(a)). Initially, the velocity of the particles relaxes as
\begin{equation}
    v(t) =
    \begin{cases}
        v_{\rm init} - t, & v_{\rm init} > t, \\
        0, & \text{otherwise,}
    \end{cases}
\end{equation}
which follows directly from Eq.~\eqref{eq:dry} in the absence of noise. Assuming a Laplacian velocity distribution \eqref{eq:Laplacian} for the initial high-temperature system, the early-time relaxation of the temperature can be estimated as
\begin{eqnarray}
\label{eq:decay}
   T(t) &=& \frac{1}{2 k_B} \!\int_{-\infty}^{\infty} dv \;v^2(t) P(v,t)
   \nonumber\\
  &\approx& \frac{1}{2 k_B \epsilon}\!\int_{t}^{\infty} dv \;(v - t)^2 e^{-v/\epsilon} = \frac{\epsilon^2}{k_B} e^{-t/\epsilon}\nonumber\\
   & = & T_{\rm init}\exp\!\left[-\frac{t}{\sqrt{k_B T_{\rm init}}}\,\right].
\end{eqnarray}
This analysis shows that the exponential decay rate of the temperature \textit{depends} on the initial temperature, in contrast to Stokes (viscous) friction, where the rate is governed solely by the viscous coefficient. The exponential decay, shown by dashed colored lines in Fig.~\ref{fig:Mpemba_temp}(a), agrees well with the simulation results in the initial stages. Consequently, initially warmer systems cool down exponentially faster than hotter ones, and therefore may \textit{overshoot} the target bath temperature (shown as a dashed horizontal line in Fig.~\ref{fig:Mpemba_temp}(a)). As a result, they require a longer time to relax to the bath temperature. This overshooting effect is absent when starting from high initial temperatures, as the corresponding fat tails in the velocity distribution shown in Fig.~\ref{fig:steady-state}(a) -- characteristic of low and high temperatures but suppressed at intermediate values -- enable a more direct relaxation pathway toward the target distribution. 

Let us now characterize the \textit{thermal Mpemba effect (TME)}, i.e.\ in terms of temperature relaxation. We emphasize that TME and EME are not identical: the distance captures the evolution of the full probability distribution, whereas the temperature reflects only its second moment, and the corresponding Mpemba effects are therefore not generally equivalent \cite{lu2017nonequilibrium, Megias/etal:2022}.\\
The relaxation time $t_{\rm relax}$ is defined as the earliest time at which the system reaches and remains within a small neighborhood of the target temperature:
\begin{subequations}
\label{eq:temp-relax}
\begin{equation}
t_{\rm relax} = \inf \Big\{ t > t_{\rm relax} \;\big|\; 
\frac{|T(t) - T_{\rm bath}|}{T_{\rm bath}} \le \delta \Big\}.
\end{equation}
where $\delta$ is defined as
\begin{equation}
\delta = \frac{|T(t_{\rm relax}) - T_{\rm bath}|}{T_{\rm bath}},
\end{equation}
\end{subequations}
and set $\delta = 0.1$. The value is chosen to be large enough to avoid transient numerical fluctuations yet small enough to indicate that the system has effectively reached the steady-state.
As shown in Fig.~\ref{fig:Mpemba_temp}(b), the Mpemba effect is also observed in terms of temperature relaxation, yielding results that closely mirror those obtained from the relaxation of the distance $\mathcal{D}$. This consistency is remarkable, as it indicates that, despite their distinct physical meanings, both measures capture the interplay between friction and activity that drives the anomalous relaxation.

\section{Conclusion}

In conclusion, the observed Mpemba effect emerges as a generic consequence of nonlinear friction dynamics, which inherently lead to the temperature overshooting during relaxation. Our results highlight the fundamental role of the interplay between friction and activity, offering different perspectives on anomalous cooling phenomena in this class of systems. The simplicity and generality of our model suggest that the Mpemba effect extends beyond conventional temperature definitions and can manifest itself in broader classes of observables which are characterized by fat tails in their probability distributions. This opens different avenues for exploring generalized Mpemba-like behavior in various physical settings, including but not limited to active particles moving through disordered environments \cite{Leyva/Pagonabarraga:2024}, relaxation processes in dense granular flows \cite{rojas2015relaxation} or in controlled protocols such as thermodynamically equidistant temperature quenches \cite{Lapolla/Godec:2022, ibanez2024heating}. In contrast to other Mpemba-like effects observed previously in granular systems \cite{Lasanta/etal:2017,santos2020mpemba, Biswas/etal:2020,mompo2021memory,Torrente/etal:2019}, the frictional Mpemba effect found here arises due to the overshooting of highly mobile particles as they relax toward the steady state. Thus, the phenomenon is amenable to experimental observation at the macroscopic scale and is of great practical significance for controlling dynamical properties across a wide range of active platforms, such as vibrobots on frictional surfaces or swarm robotics operating in complex environments \cite{Jdeed/etal:2017}.

{\it {Acknowledgment.}}
HL acknowledges the financial support by Deutsche
Forschungsgemeinschaft (German Research Foundation), Project LO 418/25-1.

\end{document}